\documentclass[acus]{JAC2000}

\usepackage{graphicx}

\begin{document}
\title{\flushright{WEBT001}\\[15pt] \centering
OPEN SOURCE REAL TIME OPERATING SYSTEMS OVERVIEW
\thanks{Thanks to Ric Claus for kindly borrowing me the MVME2306 computer}}

\author{T. Straumann, SSRL, Menlo Park, USA}

\maketitle

\begin{abstract}
Modern control systems applications are often built on top of a real
time operating system (RTOS) which provides the
necessary hardware abstraction as well as scheduling, networking
and other services. Several open source RTOS
solutions are publicly available, which is very attractive,
both from an economic (no licensing fees) as well as from a
technical (control over the source code) point of view.
This contribution gives an overview of the RTLinux and RTEMS
systems (architecture, development environment, API etc.).
Both systems feature most popular CPUs, several APIs
(including Posix), networking, portability and optional commercial
support. Some performance figures are presented, focusing on
interrupt latency and context switching delay.
\end{abstract}

\section{INTRODUCTION}

Apart from hard-real time interrupt handling and scheduling 
services, there are other OS features of interest, such as
the available APIs, target CPU architectures and BSPs (board
support packages), support for multiple processors, networking,
file systems, dynamic object loading, memory protection and so 
on.

Other important issues are licensing terms, development
environment and debugging tools and the availability of
these tools for specific host platforms.

The next section gives an overview of RTL and RTEMS
looking at some of these issues. In section 3, some performance
measurements are presented comparing the results for RTL
and RTEMS to a commercial system (vxWorks).

\section{RTLinux AND RTEMS OVERVIEW}
\subsection{RTLinux}
\subsubsection{General Information}
RTL development started at the New Mexico Institute of Mining and
Technology and is now maintained by FSMlabs Inc. which also offers
commercial support. The basic mechanism is protected by a US patent;
RTL (having a license for using the mechanism) itself is licensed
under the terms of the GPL.

RTL is distributed as a patch against certain versions of Linux
and a collection of kernel modules.

Further information about RTL is available at \cite{rtl-ref}.

\subsubsection{System Architecture}
The basic idea of RTL is striking simple: A slim layer of
software is ``hooked'' into standard Linux' interrupt handlers
and interrupt enabling/disabling primitives, thereby
effectively taking over the machine which is then managed
by a special real-time scheduler. Linux continues to run as
a low priority task.

The real-time core manages all
hardware interrupts, dispatching them appropriately, either to
Linux or to real-time threads. The interrupt manager never
allows Linux to disable interrupts. Instead, Linux disabling
an IRQ actually invokes an RTL hook which marks the target
interrupt as ``disabled''. If the interrupt manager detects 
such a marked IRQ, it holds off dispatching it to Linux until
the corresponding call to re-enable the IRQ in question is
intercepted.

Linux being a low priority task with no direct access to
the interrupt hardware implies that any real-time thread
introduced into the system can only very weakly interact 
with Linux (through special communication channels) and
may only build upon the services of the RTL core, such as
synchronization primitives and the scheduler.
Note that this low-level environment does not provide a C or ``math''
library nor any of Linux' standard system services like
networking, file systems or drivers. While the former (C library)
functionality is easy to add, providing the latter is far more complex
for obvious reasons.  Communication with user space processes
is established through special ``real-time fifo'' devices.

Following the philosophy of RTL, most of an application
should be implemented in user space, as ordinary Linux
programs. Only the real-time critical tasks go into a
special module which is loaded into kernel memory using
Linux' standard kernel-module loader.

\subsubsection{API and General Features}
Version 3.1 of RTL offers (a subset of) the POSIX (1003.13) ``pthreads'',
semaphores, condition variables and a proprietary interface to
the interrupt subsystem.
As already mentioned, no C-library is provided per-se. RTL special
features include periodic scheduling with high timing resolution.

While high timing resolution is certainly desirable, its
usefulness is reduced to some degree by the relatively high
latencies (see measurement section).

As a consequence of the layered system architecture with Linux
on top of the RTL core, an RTL application must be carefully
separated into real-time critical and non-critical parts. Only
the latter may use the powerful features of Linux, both
in kernel or user space. Critical tasks {\em must not} e.g. write
files or access non-RT drivers but they must delegate this
work to non-real time code.

\subsubsection{Supported Target Architectures}
RTL supports a subset of the CPUs and platforms supported by Linux.
x86, PowerPC, Alpha and MIPS are currently supported by RTL; at 
least on x86, SMP is supported.

\subsubsection{Development Environment}
RTL development is usually done using the well-known GNU tool chain
which has been ported to a wide variety of host platforms.
The RTL core provides support for debugging  real-time modules.

\subsection{RTEMS}
\subsubsection{General Information}
RTEMS stands for ``Real Time Executive for Multiprocessor Systems'',
where the original meaning of the letter {\em M}, namely ``Missile''
and later ``Military'' has eventually reached a civilian status.

RTEMS was developed by OaR Corp. on behalf of the US DoD and is
licensed under a GPL variant. OaR coordinates development efforts
and offers commercial support for RTEMS and other related services.

RTEMS has reached production quality and is used by military,
industrial and scientific projects. EPICS, a control systems software
which is widely used in the accelerator community, has been
ported to RTEMS as of the new 3.14 EPICS release.

More information about RTEMS can be found at \cite{rtems-ref}.

\subsubsection{System Architecture}
RTEMS was designed as a true RTOS from scratch, targeting embedded
systems, possibly with few memory. Consequently, various system
components are partitioned into separate modules (``managers'' in
RTEMS terminology) which are linked to the application
as needed. The system can further be tailored to an application's
specific needs by choosing appropriate configuration parameters.

A typical RTEMS application is built by compiling the application
itself, which must provide the necessary configuration parameters,
and linking it to the desired RTEMS managers (which are provided
in libraries) thereby creating an executable for downloading
to the target system or burning into ROM etc.

Since RTEMS is an RT system ``from the ground up'', all system
services and libraries are directly available to any application
task. 

\subsubsection{API and General Features}
RTEMS features POSIX (1003.1b), ITRON and ``classic/native'' APIs
in C and ADA (native API only) language bindings. The usual components
of an RTOS are available, such as multitasking (thread creation
and control), synchronization primitives (mutexes, semaphores,
message queues, events etc.), schedulers (fifo/round robin,
rate monotonic), clocks etc.

RTEMS provides a port of the BSD TCP/IP networking stack and supports
multiple (possibly heterogeneous) CPUs.

Like in vxWorks, no memory protection is available; the
system and application software share the same, flat memory
space.

RTEMS itself does not ship a shell as powerful as vxWorks'
nor does it offer a dynamic loader.
However, there are ongoing efforts of creating application
programs providing the respective features.

The only file systems currently implemented are a
remote TFTP and a ``in memory'' (ramdisk) file system.

\subsubsection{Supported Target Architectures}
RTEMS is designed to be easily portable and consequently it
supports many CPU architectures, such as m68k, ColdFire,
Hitachi SH, Intel i386, Intel i960, MIPS, PowerPC, SPARC, AMD A29k
and HP PA-RISC.

\subsubsection{Development Environment}
RTEMS uses the GNU tool chain.

%\newlength{\cw}\setlength{\cw}{0.2\columnwidth}
%\newcommand{\ff}{\footnotemark[2]}

%\begin{table}[h]
%\begin{minipage}{\columnwidth}
%\hspace*{\fill}
%\begin{tabular}{p{\cw}|p{\cw}|p{\cw}|p{\cw}}
%	& RTL & RTEMS & vxWorks \\
%\hline
%License		& GNU & GNU & commercial\\
%\hline
%RT API  & POSIX & POSIX, ITRON, native & native, POSIX \footnotemark[1] \\
%\hline
%Networking & y\ff & y & y\\
%\hline
%Multiple CPUs & y & y & y\\
%\hline
%Host Tools   & GNU & GNU & GNU/Tornado\\
%\hline
%File Systems  & NFS\ff, Ramdisk\ff, FAT\ff, others\ff & TFTP, Ramdisk & NFS, TFTP, FAT, others \\
%\hline
%Shell         & y\ff & y\footnotemark[3] & y\\
%\hline
%Dynamic Loader & y & n & y\\
%\hline
%Targets		& \multicolumn{3}{c}{see text}\\
%\hline		
%Host Arch.	& many\footnotemark[4] & many\footnotemark[4] & Win, Solaris, HPUX\\
%\hline
%\end{tabular}
%\hspace*{\fill}
%\end{minipage}
%\caption{Comparison of some RTOS key features}
%\label{tab-compar}
%\end{table}

\section{RESPONSE TIME PERFORMANCE TEST}
A key property of any hard-real time system is its ``response time'',
i.e. the time it takes for the system to react to some external
event under worst case conditions. Two important terms shall be
defined here:
\begin{Description}
\item[``Interrupt Latency''] The time it takes from a device asserting
an interrupt line until the system dispatching the
corresponding interrupt handler (ISR) shall be called {\em interrupt latency}.
\item[``Context Switch Delay''] This term defines the time it takes
to schedule a task. It involves the scheduler determining which task
to run, saving the current task context and restoring the new one.
\end{Description}
Of course, it is practically impossible to find the worst case
conditions given the huge number of possible state combinations that
can occur in a computer system.

Therefore, a statistical approach is taken to create ``worst case''
conditions. The idea is to let the system operate under heavy load
for some time while measuring the latencies. The maximal delay recorded
during the test is then assumed to reflect the ``worst case''.

\subsection{Test Algorithm}
A PowerPC 604 CPU (300MHz) on a MVME2306, PReP compatible
board by Motorola was chosen to perform the measurements. BSPs for RTL, RTEMS
and VxWorks were available, allowing for comparison of the three
systems on the same target hardware.

The MVME2306 (like most PPC platforms) features timer hardware
with a reasonable resolution, which can be set up to generate
periodic interrupts. Because the running timer is readable
``on-the fly'', a precise measurement of latencies can easily
be accomplished.

The test software package \cite{soft-ref}
consists of an initialization routine,
an interrupt handler (ISR) and a simple ``measurement'' procedure.

The initialization code sets up the timer hardware, connects
the ISR to the respective interrupt and spawns a task (MT) executing
the measurement procedure at the highest priority available on the
system under test.

The ISR determines the interrupt latency by reading the timer
and notifies the MT by releasing a semaphore on which the MT
blocks. This causes the system to schedule the MT (having become
the highest priority runnable task), which,
reading the running timer is able to determine the time that
elapsed from the ISR releasing the semaphore until the MT
actually getting hold of the CPU. After recording the delay,
the MT again blocks on the semaphore.

This simple test was performed on a system heavily
loaded with low priority tasks, networking and serial
I/O traffic causing a large volume of interrupts (also
at a priority lower than the timer hardware IRQ).

According to the definition, a hard-real time system
must guarantee that the latencies experienced by the high priority
ISR and MT stay below a certain {\em hard} limit, regardless
of the amount of low-priority load (note that interrupts inherently
have a higher priority than any normal task, hence a low-priority
interrupt still interrupts a high-priority task).

Hence, the maximal recorded latencies during the test
constitute a measure for the quality of a given system.

\subsection{Results}
The test was performed on the same hardware under the
RTL, RTEMS and VxWorks systems. 2'000'000 timer interrupts
were generated at a rate of 4kHz and the
maximal and average latencies were recorded. Measurements
were made under both, idle and loaded conditions.

The load that was imposed on the system under test
consisted of ``flood pinging'' its network interface from a host
computer, while letting a low priority thread copy characters
from a TCP socket (connected to the host's ``chargen'' port)
to the serial (RS-232) console. 
Thus, the loaded system was subject to heavy interrupt
and kernel activity involving scheduling, synchronization
primitives, networking and driver code sections among others.

The results are shown in Tab.~\ref{tab-results}.
The idle systems all exhibit comparable figures. The situation
changes, however, quite dramatically under load:
Whereas RTEMS and VxWorks show similar performance,
RTL's latencies are substantially higher on the loaded system.
This is not really surprising given the far more complex
interrupt dispatching that is needed to manage and emulate
the Linux interrupts.

\begin{table}
\begin{minipage}{\columnwidth}
\hspace*{\fill}
\begin{tabular}{@{}r|r@{.}lr@{.}l@{$\pm$}r@{.}lr@{.}lr@{.}l@{$\pm$}r@{.}l@{}}
%vxWorks	& 13&1 & (1&98 & 0&16)	 & 19&0 & 3&06 & 0&28 	& native threads, native semaphores\\
%RTL	& 13&5 & (1&69 & 0&15)	 & 33&1 & 8&72 & 0&52 	& pthreads, psemas\\
%RTEMS	& 14&9 & (1&32 & 0&06)	 & 16&9 & 2&29 & 0&13	& pthreads, psemas\\
%RTEMS	& 15&0 & (1&33 & 0&11)	 & 27&0 & 2&28 & 0&15	& native threads, psemas\\
%RTEMS	& 15&1 & (1&32 & 0&08)	 & 16&4 & 2&16 & 0&14	& native threads, events\\
%vxWorks	& 25&2 & (2&87 & 1&47)	& 38&8  & 9&51 & 3&20	& native threads, native semaphores\\
%RTL	& 196&8& (2&12 & 3&32)	& 193&9 & 11&15 & 4&5	& pthreads, psemas\\
%RTEMS	& 19&2 & (2&41 & 1&69)	& 213&0 & 10&44 & 12&70	& pthreads, psemas\\
%RTEMS	& 15&0 & (2&69 & 1&52)	& 49&1  & 5&32 & 2&67	& native threads, psemas\\
%RTEMS	& 20&5 & (2&91 & 1&76)	& 51&3  & 3&72 & 2&02	& native threads, events\\
%%%% rounded:
	& \multicolumn{6}{c}{Interrupt Latency}
	& \multicolumn{6}{c}{Context Switching}
\\
	& \multicolumn{2}{c}{max} & \multicolumn{4}{c}{avg$\pm\sigma$}
	& \multicolumn{2}{c}{max} & \multicolumn{4}{c}{avg$\pm\sigma$}
\\
\hline
\multicolumn{1}{c}{} & \multicolumn{12}{c}{\em Idle System} \\
RTL	& 13&5 & (1&7 & 0&2)	& 33&1 & (8&7 & 0&5) 		\\
RTEMS\footnotemark[1]	& 14&9 & (1&3 & 0&1)	& 16&9 & (2&3 & 0&1)		\\
%RTEMS 	& 15&0 & (1&3 & 0&1)	& 27&0 & (2&3 & 0&2)		\\
RTEMS	& 15&1 & (1&3 & 0&1)	& 16&4 & (2&2 & 0&1)		\\
vxWorks & 13&1 & (2&0 & 0&2)	& 19&0 & (3&1 & 0&3)	\\
\multicolumn{1}{c}{} & \multicolumn{12}{c}{\em Loaded System} \\
RTL	& 196&8& (2&1 & 3&3)	& 193&9 & (11&2 & 4&5)		\\
RTEMS\footnotemark[1]	& 19&2 & (2&4 & 1&7)	& 213&0 & (10&4 & 12&7)		\\
%RTEMS 	& 15&0 & (2&7 & 1&5)	& 49&1  & (5&3 & 2&8)		\\
RTEMS	& 20&5 & (2&9 & 1&8)	& 51&3  & (3&7 & 2&0)		\\
vxWorks	& 25&2 & (2&9 & 1&5)	& 38&8  & (9&5 & 3&2)		\\
\end{tabular}
\hspace*{\fill}
% arrgh; need a hack for this footnote...
\footnotetext{\footnotemark[1]using pthreads}
\end{minipage}
\caption{Latency measurement results. All times are in $\mu$s.
vxWorks and RTEMS use native threads unless otherwise noted.
RTL uses the pthread API.}
\label{tab-results}
\end{table}

Somewhat surprising is RTEMS' increased scheduling latency
when using the pthread API, as one would assume the
implementation to merely consist of an inexpensive wrapper 
to the native API. Given the good performance of
the latter, one can expect however, that making improvements
should be relatively straightforward.

As can be seen, the average latencies are about an order of magnitude
less than the respective maxima. Although our statistical test
gives some lower bound of the maximal latencies, it is impossible
to draw conclusions about the true worst case figures which are
obviously extremely difficult to establish.

Usually, the interrupt handling parts of any system are
highly hardware-architecture dependent. Therefore, while
representative for the PowerPC, the interrupt latency
figures stated here can not easily be generalized to other
CPU architectures.

\section{CONCLUSION}
RTEMS and RTL are two quite different open-source RTOS
solutions.

RTEMS seems to offer both, core features and performance
which are comparable to a commercial system like vxWorks.

RTL could be interesting in situations, where the full
power of a desktop system is needed, enhancing such a
system by hard-real time features. This comes, however,
at the expense of higher latencies (compared to RTEMS or vxWorks)
and limitations of system services that are available to
the real-time tasks.

Finally, it should be noted, that the simple benchmark
presented in this paper does by no means constitute a
thorough performance evaluation and comparison, an arduous
task to which the interested reader is encouraged to contribute.

\end{document}